\documentclass[%
 aip,
 jmp,%
 amsmath,amssymb,
preprint,%
]{revtex4}

\draft 

\usepackage{graphicx}
\usepackage{dcolumn}
\usepackage{bm}

\begin{document}





\title{Magneto-Plasmonics and Resonant Interaction of Light with Dynamic Magnetisation in Metallic and All-Dielectric Nanostructures (Review)}
\author{Ivan S. Maksymov}
\email{ivan.maksymov@uwa.edu.au}
\affiliation{School of Physics, M013, University of Western Australia, Crawley, 6009, WA, Australia}

\date{\today}

\begin{abstract}
A significant interest in combining plasmonics and magnetism at the nanoscale gains momentum in both photonics and magnetism sectors that are concerned with the resonant enhancement of light-magnetic-matter interaction in nanostructures. These efforts result in a considerable amount of literature, which is difficult to collect and digest in limited time. Furthermore, there is insufficient exchange of results between the two research sectors. Consequently, the goal of this review paper is to bridge this gap by presenting an overview of recent progress in the field of magneto-plasmonics from two different points of view: magneto-plasmonics, and magnonics and magnetisation dynamics. It is expected that this presentation style will make this review paper of particular interest to both general physical audience and specialists conducting research on photonics, plasmonics, Brillouin light scattering spectroscopy of magnetic nanostructures and magneto-optical Kerr effect magnetometry, as well as ultrafast all-optical and THz-wave excitation of spin waves. Moreover, readers interested in a new, rapidly emerging field of all-dielectric nanophotonics will find a section about all-magneto-dielectric nanostructures.
\end{abstract}


\maketitle 

\section{Introduction}

Light technology has revolutionised society through its applications in medicine, communications, entertainment and culture. Nowadays, industries based on light are major techno-economic drivers, enabling cutting-edge technologies such as solar panels (see, e.g., the introduction to a focus issue on photovolatics \cite{solar_cells_editorial}), LED lighting \cite{Nobel2014}, computer monitors and mobile phone screens, digital cameras and projectors \cite{Aos10}, as well as optical tomography and intravascular photoacoustic imaging \cite{Jan14}. These and many other devices give us better access to information, more reliable health care, better ways of saving energy and new forms of entertainment. To highlight the important role of light-based technologies in our society, the year $2015$ was proclaimed by UNESCO to be the International Year of Light and Light-based Technologies. 

It is noteworthy that the research and commercial potentials of light-based technologies are much larger than the state-of-the-art of science and technology and the current market, respectively. For example, microelectronic chips made from silicon are cheap and easy to mass-fabricate, and can be densely integrated. Consequently, researchers in the photonics sector try to create silicon-based optical devices that can exploit the benefit of silicon while also being fully compatible with electronics (see the introduction to a focus issue on silicon photonics \cite{Editorial} and, e.g., Refs. \cite{Alm04,Vivien_Pavesi}). 

The development of new photonic technologies is very important because semiconductor microelectronics is expected to reach the fundamental limits of its functionality in the observable future. These limits are often discussed in the context of the Moore's law \cite{Sch97} and its limitations \cite{Zhi03}. In particular, it is foreseen that integrated photonic circuits will allow optical systems to be more compact than discrete optical components and to have higher performance as compared with microelectronic circuits. Integrated photonic circuits are also expected to be compatible with nanoelectronic components \cite{Lu07} and to be better protected from electromagnetic interference. Finally, it is envisioned that integrated photonic circuits will be capable of exploiting the fundamental aspects of the quantum nature of light and matter, and to develop applications relating to quantum information and quantum computing, as well as measurements at and beyond the standard quantum limits \cite{Wal09,Xia13}.    

However, it is unknown how fast integrated photonic circuits or at least their individual components (e.g., optical interconnects \cite{Sav02,Kin15}) can complement and/or substitute semiconductor microelectronic chips. In practice, an integrated photonic circuit has to bring together multiple photonic functions \cite{Wil14} on a single chip. In order to achieve this goal one has to have reliable mechanisms of ultrafast all-optical signal processing at the micro- and nanoscale, as well as suitable materials and fabrication technologies. However, whereas many existing photonic technologies are excellent for, e.g., ultrafast all-optical signal processing \cite{Wil14} and connection to the outside of the integrated device \cite{And12}, no one photonic technology can be used to realise all functions of electronics, microwave and magnetic devices as well as radio frequency antennas.

Consequently, experts agree that integrated devices of the future, which are expected to have better performance and offer previously unavailable nonclassical (quantum) functionality, will most probably rely on employing different systems (see, e.g., Refs. \cite{Wal09,Vivien_Pavesi,Xia13}). In turn, each subsystem of these systems will be specialised on fulfilling distinct tasks, and the physics of operation of each subsystem will be different.

Microwave magnetic nanodevices \cite{Kru10,Ser10,Gub10,Len11,Sta14,Hir14,Kra14,Magnonics_book} are often viewed as the most probable candidates for the integration with photonic devices. This is because many processes involved in physics of magneto-optical phenomena are well understood (see, e.g., Ref. \cite{Zvezdin_Kotov}), and many magneto-optical devices \cite{Zvezdin_Kotov,Sta09} (e.g., magneto-optical drives) had huge commercial success in the past and remain in use nowadays. Moreover, in magnetic data storage media, magnetic nanostructures have already been combined with nanoelectronics (e.g. in read heads and magnetic random access memories).

The aim of this review paper is to critically evaluate the recent progress in the integration of nanophotonic and nanomagnetic technologies. Research on this topic is being currently conducted by both photonics and magnetism sectors, which results in huge amount of literature but also leads to insufficient exchange of results between the two research communities. Consequently, this review paper also aims to bridge this gap. Of course, this review cannot cover the entire field. Several highly relevant areas have been selected without attempting to provide a full survey.


\section{Magneto-Plasmonics}

Surface plasmons are waves that propagate along a metal-dielectric interface \cite{Bar03,Sto11,Zay13}. A localised surface plasmon is the result of the confinement of a surface plasmon in a metallic nanoparticle of size comparable to or smaller than the wavelength of light used to excite the plasmon. Metallic nanoparticles at or near their plasmonic resonance generate highly localised electric field intensities that effectively correspond to light stopping at their vicinity. Varieties of nanoparticles and their constellations (e.g., nanoantennas \cite{Kra13}) were shown to enhance the local field and thus improve light-matter interaction.

Since all metals absorb light in the visible and infrared spectral ranges, maximum efficiency of a plasmonic device can be achieved by using metals with the lowest absorption cross-section. Therefore, most of the plasmonic devices are made of gold or silver because these two metals exhibit the lowest absorption losses at optical frequencies.

However, in many practical cases, these metals must be combined with optically active materials in order to provide active control of plasmons \cite{Sto11,Zay13}. In particular, plasmons can be controlled by a magnetic field applied to a hybrid device consisting of a plasmonic metal nanostructure combined with a ferromagnetic layer. While the former supports propagating or localised plasmon modes with the lowest possible absorption losses, the latter exhibits a large magneto-optical activity that opens up routes for ultrafast control of light such as, e.g., magneto-plasmonic switching or high-sensitivity biosensing.

The research field that combines magnetic and photonic functionalities is called magneto-plasmonics. Controlling the optical properties of surface plasmons using magnetic effects was probably first suggested by Chiu and Quinn \cite{Chi72}, who investigated the effect of the external static magnetic field on the dispersion relation of surface plasmons in a metal.  The review papers by Armelles et. al. \cite{Arm13} and Temnov \cite{Tem12}, a special issue on magneto-plasmonics \cite{Arm14} as well as the book \cite{Ino13} laid out the basics of modern magneto-plasmonics. The paper by Armelles et. al. \cite{Arm13} also provides an excellent introduction into the magneto-optical Faraday and Kerr effects. Consequently, in this section we focus on developments that have emerged since those works appeared. We also analyse results presented in earlier works that were not discussed in detail in the previous review papers, but are of immediate relevance to this review paper.

\subsection{Plasmon-Enhanced Transverse Magneto-Optical Kerr Effect}

As shown in Refs. \cite{Arm13,Tem12}, many works on magneto-plasmonic nanostructures demonstrate the possibility to enhance the magneto-optical effects by exploiting resonance properties of surface plasmons in gratings, which combine a nanopatterned thin layer of a noble metal (gold) with magneto-insulating thin film (bismuth iron garnet, BIG) \cite{Bel11}. Such structures offer a combination of a large Faraday rotation (owing to the BIG film) and small optical losses for wavelengths longer than $650$ nm (owing to the nanostructured noble metal). The cross-polarised transmission and polar Kerr rotation in similar structures were measured in separate works as a function of external static magnetic field \cite{Wur08}. However, enhancement of magneto-optical effects by means of surface plasmons was not demonstrated. Previous theoretical works have also predicted that the Faraday and Kerr effects can be resonantly increased in hybrid gold-BIG structures, in particular near the Wood's anomalies \cite{Bel07,Bel09}. These predictions have been confirmed in Ref. \cite{Bel11} by observing significant enhancement of the transverse magneto-optical Kerr effect (TMOKE) in transmission.

The TMOKE is defined as a change of reflected intensity of \textit{p}-polarised light when the direction of the external static magnetic field is changed from the saturated state $+M_{\rm{s}}$ to $-M_{\rm{s}}$, being $M_{\rm{s}}$ the saturation magnetisation \cite{Zvezdin_Kotov}. However, as already mentioned above, the TMOKE signal can also be detected in transmission by replacing the reflection coefficient by the transmission coefficient. The attainment of a large TMOKE response is important for many practical applications that include, but not limited to, 3D imaging \cite{Aos10}, magnonics (see Section $3$), magnetically tuneable optical metamaterials, and magneto-optical data storage \cite{Zvezdin_Kotov}.

Plasmon-enhanced TMOKE in a planar waveguiding configuration may also be useful for miniaturised photonic circuits and switches, which can be controlled by external magnetic fields \cite{Kre13}. The cited paper experimentally demonstrates a new concept for the magnetic  modulation of the light transmission through at a high level of optical transparency. This concept incorporates magneto-insulating thin films with waveguide-plasmon polaritons \cite{Chr04}, which stem from the coupling between localised surface plasmons and guided optical modes. The TMOKE signal is quantified as the relative change in the intensity of the transmitted light when the magnetisation is reversed. A large TMOKE signal of and a high experimental transmittance of $45\%$ were demonstrated. However, as will be shown below, this value must be increased in order to make the observed effect more suitable for practical applications.

\subsection{Plasmon-Enhanced Faraday Effect}

Relative change in the intensity of the transmitted light due to the TMOKE is also known as the "transverse Faraday effect" \cite{Man99}. However, more often under the Faraday effect (or Faraday rotation) one understands a rotation of the plane of polarisation which is proportional to the component of the magnetic field in the direction of propagation \cite{Arm13}. Apart from a few applications in measuring instruments, sensing, and spintronics, the Faraday rotation can be used for amplitude modulation of light, and are the basis of optical isolators and optical circulators, which are required in optical telecommunications and other laser applications.

It is noteworthy that dimensions of miniaturised commercial optical isolators are about the size of the tip of a pencil. These miniaturised devices offer opportunities for laser diode manufacturers requiring an in-package ultra-high performance isolator. However, their dimensions are not suitable for the application in integrated photonic circuits.

Consequently, one has to utilise arrays of metallic nanoparticles or metallic gratings combined with magneto-insulating films made from, e.g., BIG. The first approach have been used in Ref. \cite{Pra09}, in which enhanced optical Faraday rotation has been reported in gold-coated maghemite nanoparticles. However, the enhancement of the Faraday rotation achieved using a single nanoparticle may be insufficient for practical applications. 

The second approach relying on the combination of metallic gratings with BIG films have been used in Ref. \cite{Chi13} [Fig. $1$(a-c)]. The BIG film was prepared by pulsed laser deposition (see, e.g., Ref. \cite{Weh11}). First, a thin buffering layer of $10$-nm yttrium iron garnet (YIG) was deposited on a glass substrate and was annealed at 1000$^{\rm{o}}$. After that, another layer of $140$ nm BIG was deposited on the YIG buffering layer. As the next step, gold nanowires were fabricated on the BIG film by electron beam lithography and a subsequent lift-off process.

The fabricated system allows one to simultaneously excite plasmonic and photonic modes of the hybrid structure. A localised plasmon resonance is excited by incoming light polarised perpendicular to the wires. Moreover, the nanowires introduce periodicity to the hybrid structure and enable light to couple into the thin BIG film. Consequently, the nanostructure acts as a planar photonic crystal waveguide for photonic modes \cite{Chr04}. The localised plasmon resonance and the waveguide resonance interact strongly, and this interaction can be controlled by applying external static magnetic field. As a result, Faraday rotation has been increased by up to $\sim9$ times compared with the bare BIG film, while high transparency is maintained [Fig. $1$(d,e)].

\begin{figure}[htbp]
\centering\includegraphics[width=12cm]{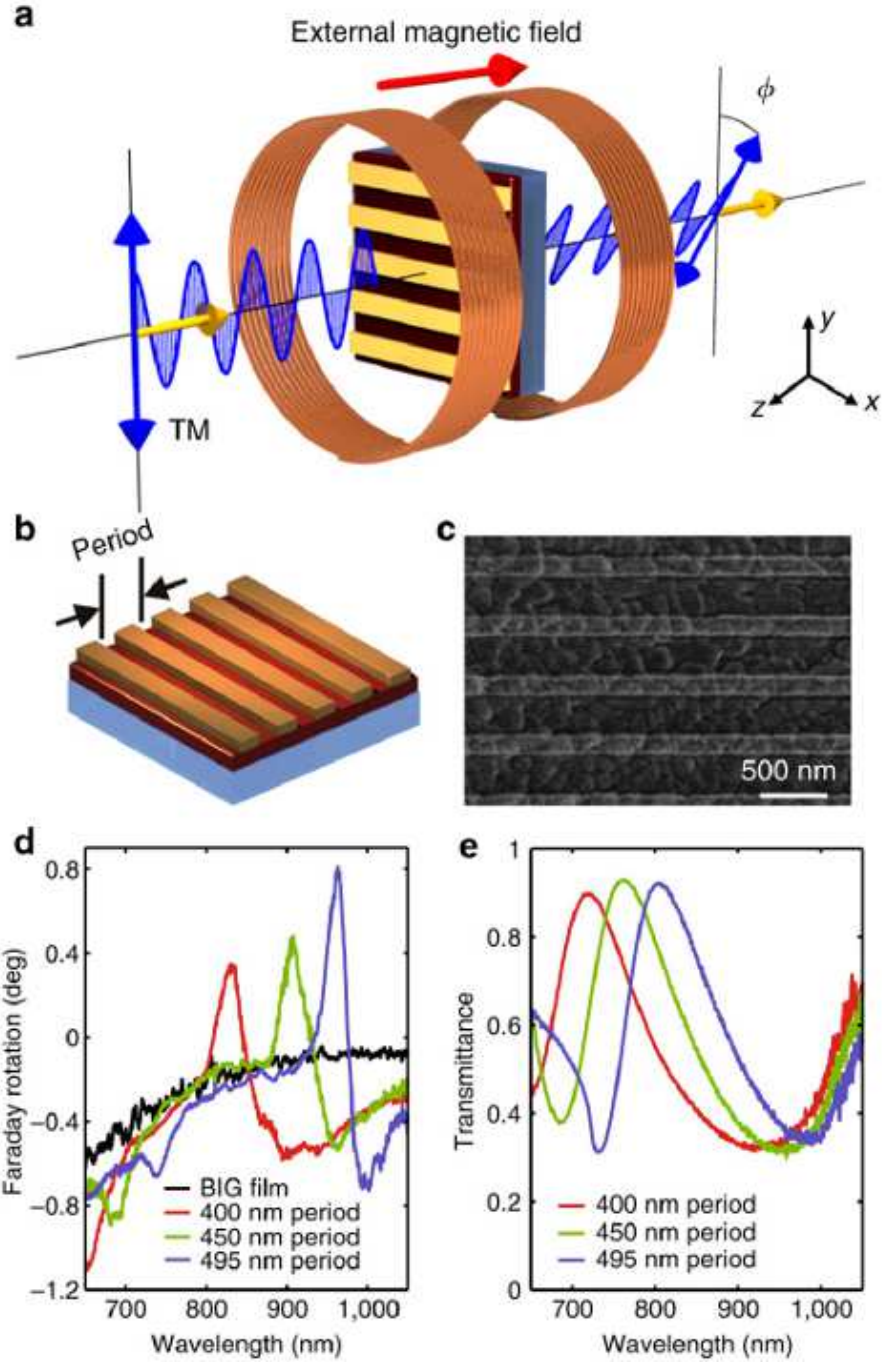}
\caption{(a) Faraday rotation by a magneto-plasmonic photonic crystal for the TM-polarised incident light, where $\phi$ is the Faraday rotation angle. At normal incidence, TM-polarised light has the electric field perpendicular to the gold wires. (b) Schematic of the magneto-optical photonic crystal, where the BIG film (dark red) is deposited on a glass substrate (blue) and the periodic gold nanowire structure is sitting atop. (c) A scanning electron microscopy image of one of the investigated hybrid structures. (d) Measured Faraday rotation of the three samples at normal incidence (TM polarization), compared with measured Faraday rotation of the bare BIG film. (e) Measured transmittance of the three samples at normal incidence (TM polarization). Reproduced with permission from Ref. \cite{Chi13}.}
\end{figure}

We would like to discuss the problem of achieving simultaneously high Faraday effect (and also TMOKE, see Ref. \cite{Kre13}) and high transparency of the hybrid plasmonic-magneto-optic nanostructure. As shown in Fig. $1$(d-e), the spectra of Faraday rotation display a resonant feature. As the period increases, the resonance shifts to longer wavelength and the maximum Faraday rotation increases. The sample with the $495$ nm period produces a maximum Faraday rotation at $963$ nm, which is an $\sim9$-fold enhancement as compared with the Faraday rotation in the BIG film without gold wires. However, the same sample shows just $36\%$ transmittance at $963$ nm. Despite a considerable enhancement of the Faraday rotation and impressive miniaturisation of the device, this value of transmission is far from $>80-90\%$, which is required for practical applications.

The same problem of low transmission that accompanies the plasmon-enhanced TMOKE response also exists in all-ferromagnetic nanostructures, which are discussed in Section $2.3$. The explanation of the origin of this problem in the framework of the Fano resonance \cite{Mir10} and possible ways of its solution will also be summarised in Section $2.3$. 

\subsection{Plasmon-Enhanced TMOKE in All-Ferromagnetic Nanostructures}

As discussed above, microwave magnetic devices are often viewed as the most probable candidates for the integration with photonic devices. Magnonic crystals, which are an important research direction in a broader field of microwave magnetism \cite{Kru10,Ser10,Gub10,Len11,Sta14,Kra14}, are one- or two-dimensional nanostructures consisting, respectively, of ferromagnetic metal nanostripes or dots, anti-dots (holes in ferromagnetic films) and their combinations. Also, all-ferromagnetic nanostructures are potential candidates for ultra-high-density magnetic recording and they are controllable by light (see, e.g., Ref.\cite{Lam14}). 

\begin{figure}[htbp]
\centering\includegraphics[width=8.5cm]{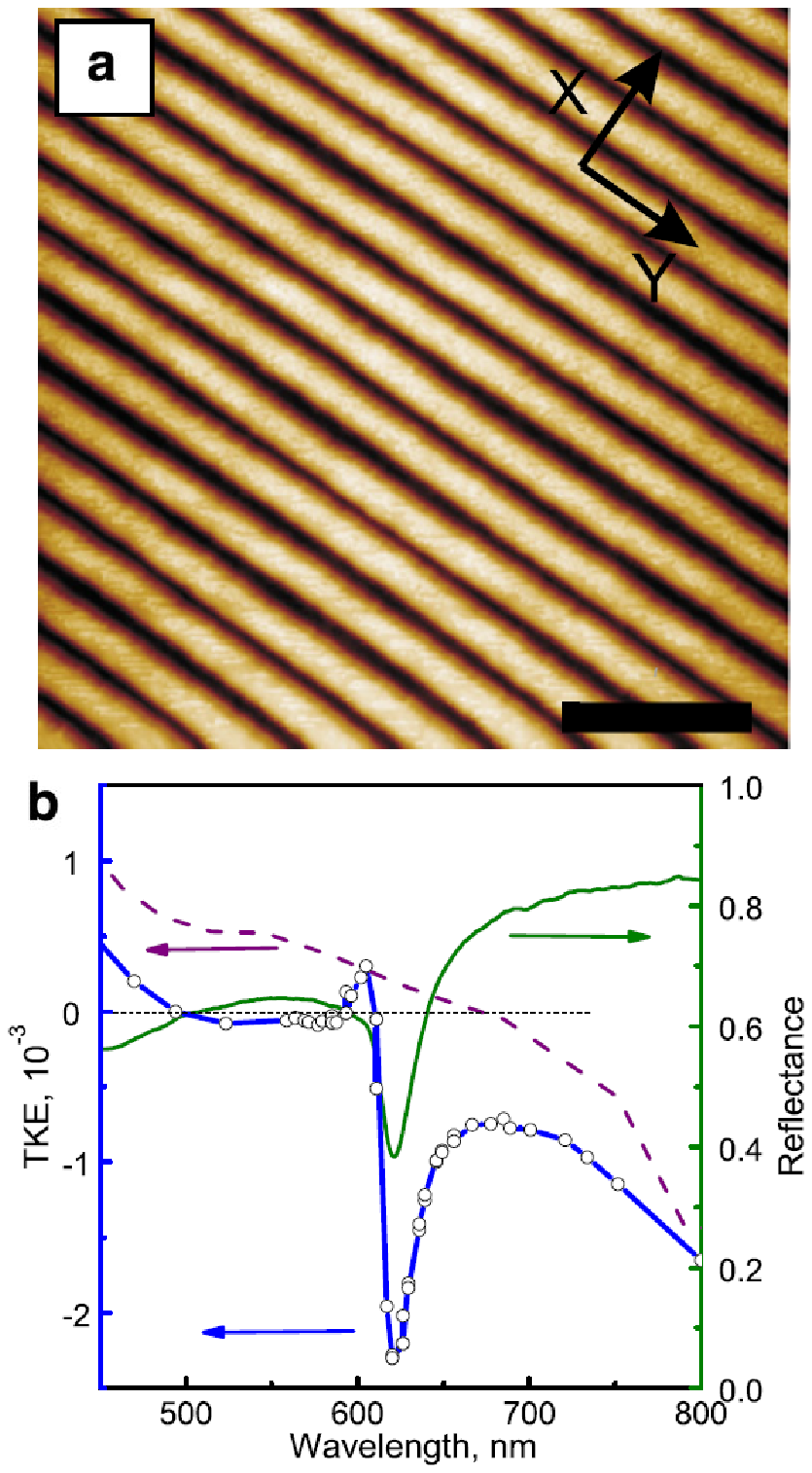}
\caption{(a) The AFM image of the all-nickel magneto-plasmonic grating. Black bar is equal to $1$ $\mu$m. (b) TMOKE spectra of the one-dimensional magneto-plasmonic grating (circles) and the reference plain nickel film (dashed curve). The solid line shows the reflection spectrum of the grating. Reproduced with permission from Ref. \cite{Gru10}.}
\end{figure}

Magnonic crystal can also be made from magneto-insulating materials such yttrium iron garnet (YIG) \cite{Ser10,Mru14_1}. Similar to BIG, YIG possesses good optical properties, which makes it possible to develop devices exploiting the physics of light-spin wave interaction \cite{Sta09}. The progress in this direction is supported by advances in the fabrication of magneto-optical components compatible with silicon platform \cite{Bi10}. Resonant optical properties of all-magneto-dielectric nanostructures will be discussed in Section $4$. 

Apart from their importance in magnonics, all-ferromagnetic nanostructures have attracted significant attention because they can be used to enhance magneto-optical effects (see, e.g., Refs. \cite{New08,Cti09,Gru10,Sap11,Bon11,Che11,Che12,NKos13,Mac13}). Indeed, although ferromagnetic metals exhibit a stronger optical damping as compared with gold or silver \cite{Che11}, resonance excitation of surface plasmons in ferromagnetic nanostructures is significant, which makes it possible to enhance the TMOKE response. 

For example, pure nickel subwavelength one-dimensional gratings [Fig. $2$(a)] without any noble-metal film inclusions \cite{Gru10} support the excitation of surface plasmons strong enough to achieve the TMOKE enhancement comparable with the case of trilayer Au/Co/Au films \cite{Arm13} and ten times larger than in a reference continuous nickel film [Fig. $2$(b)]. However, similar to the hybrid nanostructure used to enhance the Faraday effect (Fig. $1$), the maximum of the TMOKE signal produced by the grating is accompanied by a significant decrease in reflection [the green line in Fig. $2$(b)].

We would like to note that the full width of plasmonic resonances in noble-metal gratings is mostly defined by the radiative damping of surface plasmons, which leads to the Fano-type shape of the resonances that appear in the spectral vicinity of the Wood's anomaly \cite{Hes65}. It is well-known that in plasmonic nanostructures the lineshape of the Fano resonance can be manipulated by using different approaches (see, e.g., Refs. \cite{Gal09,Gal11,Mak11}).

The same strategy works when applied to all-ferromagnetic magneto-plasmonic nanostructures. For example, in Ref. \cite{NKos13} a strong plasmon-assisted TMOKE response was achieved in all-Permalloy one-dimensional grating simultaneously with the maximum of reflectivity attainable in this nanostructure (Fig. $3$). Although the maximum attainable reflectivity is just $20\%$, which is because this grating is in fact a one-dimensional magnonic crystal that was designed without optimising the optical properties, the work Ref. \cite{NKos13} clearly demonstrates a correlation between the TMOKE enhancement and the Fano resonance \cite{Mir10}. Indeed, as shown by the vertical lines in Fig. $3$(b-e), both experiment and theory demonstrate that the maximum of the plasmon-enhanced TMOKE response has been achieved at the frequency of the Fano resonance, which for the given geometry of the Permalloy grating also corresponds to nearly the maximum of the grating reflectivity.

Of course, in magneto-plasmonic experiments the Fano resonance frequency depends on the applied static magnetic field. However, in many cases experimental conditions do not allow one to measure the absolute values of reflectivity for the two opposite directions
of the magnetisation vector with accuracy which is sufficient for reliable extraction of the value of the shift in the Fano resonance frequency upon reversal of magnetisation. In Ref. \cite{NKos13}, using the simulation data this shift was estimated to be of at least $1.5$ nm. Therefore, in Ref. \cite{NKos13} a model of general asymmetric Fano resonance profile with magnified characteristic features was used to demonstrate that the maximum of the TMOKE occurs at the frequency at which the slope of the Fano resonance peak is maximum. This finding is consistent with a previous theoretical model proposed by Belotelov et. al. \cite{Bel11}.

To conclude the section on all-ferromagnetic nanostructures, we would like discuss the material of the grating (magnonic crystal) shown in Fig. $3$(a) - Permalloy (Ni$_{80}$Fe$_{20}$). As discussed in Refs. \cite{Kru10,Ser10,Gub10,Len11,Sta14,Hir14,Kra14,Magnonics_book}, Permalloy is paramount for all applications in microwave signal processing, magnetic memory, logics, and sensors. This is because of the optimum combination of microwave magnetic properties of Permalloy: the vanishing magnetic anisotropy and one of the smallest magnetic (Gilbert) damping among ferromagnetic metals. Similar to nickel that has been proven to support plasmonic resonances \cite{Gru10,Che11}, Permalloy also possesses high magneto-optical properties as confirmed by experiments on thin Permalloy films \cite{Tan63,Rob63,Kri85}. However, thinking of application in spintronics and magnonics, nickel is characterised by large magnetic losses and is known as a material with very large magnetostriction. Furthermore, its saturation magnetisation ($6000/4\pi$ Oe) is small (almost two times smaller
than for Permalloy and almost four times smaller than for iron). For all these reasons, use of nickel in spintronics and magnetisation dynamics is limited.

\begin{figure}[htbp]
\centering\includegraphics[width=16cm]{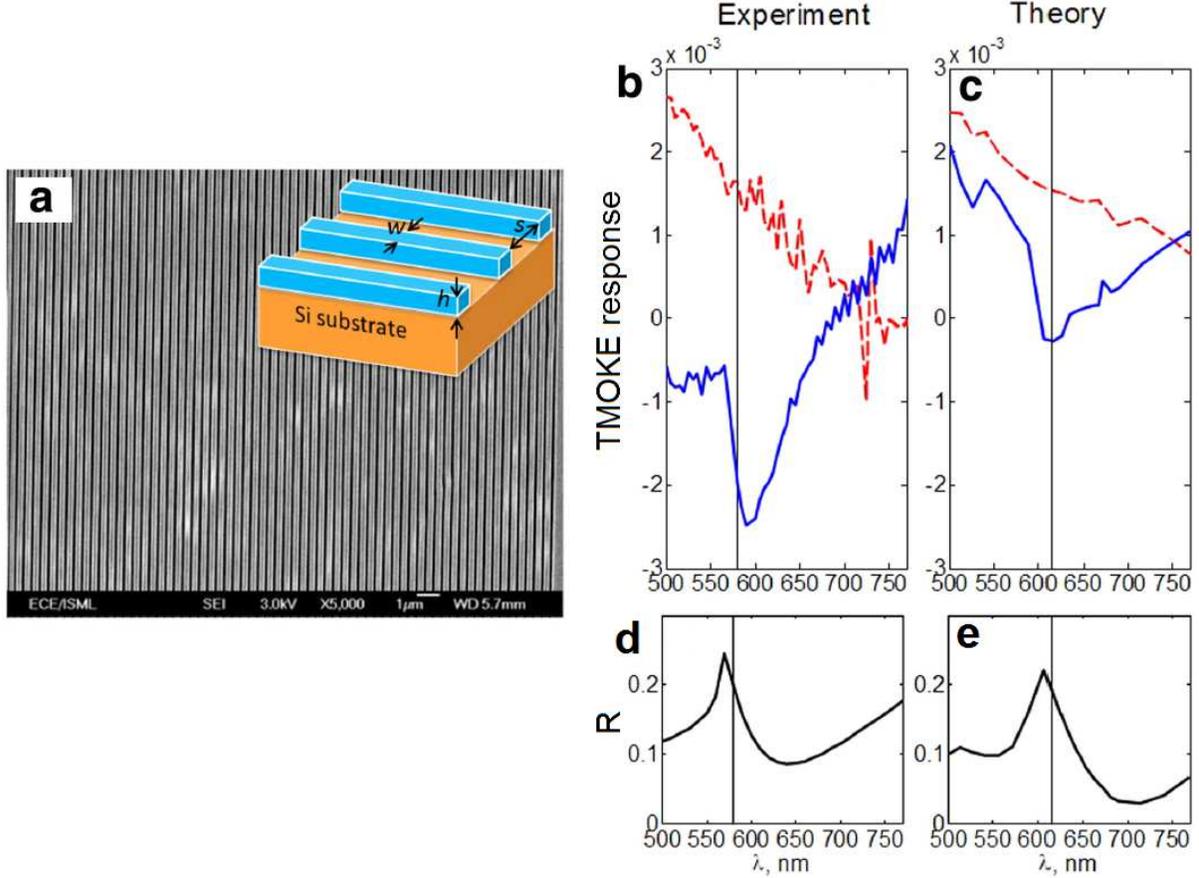}
\caption{(a) Scanning electron micrograph and schematic of the Permalloy grating (one-dimensional magnonic crystal). The total area of the grating is $0.5 \times 0.5$ cm$^2$, $h=100$ nm, $w=264$ nm, and $s=113$ nm. The thickness of the Si substrate is $0.8$ mm. Measured (b) and simulated (c) TMOKE response of the $100$ nm-thick reference Permalloy film (dashed line) and Permalloy grating (solid line). Measured (d) and simulated (e) reflectivity spectra of the Permalloy grating. The vertical straight solid line denotes the Fano resonance wavelengths.  Reproduced with permission from Ref. \cite{NKos13}.}
\end{figure}

\subsection{Longitudinal Magneto-Photonic Intensity Effect}

The Faraday and Kerr effects have been known for more than 150 years and they, as well as their different configurations \cite{Zvezdin_Kotov,Arm13}, have been exploited in various devices. However, it turns out that a combination of these effects with unique optical properties of high-quality nanostructures may lead to the discovery of new magneto-optical effects, which open up novel opportunities to control light with magnetic fields at the nanoscale. 

Belotelov et. al. \cite{Bel13_LMPIE} investigated a nanostructure consisting of three layers: a non-magnetic dielectric substrate, followed by a magneto-insulating layer, and a thin gold film sitting on top. The gold layer was periodically perforated by parallel slits with period $d=661$ nm. The lower refractive index of the substrate relative to one of the magnetic layer ensured the existence of guided optical modes in the magnetic layer.

In absence of an external magnetic field, the magnetic layer is demagnetised and the investigated structure supports surface plasmons and waveguided transverse magnetic (TM) or transverse electric (TE) waves. Simulations demonstrated that the TM and TE modes are mostly localised in the magnetic layer. Moreover, it was shown that these modes are sensitive to the permittivity of the adjacent metal and therefore also have plasmonic character.

When an external static magnetic field is applied, the presence of non-diagonal terms of the dielectric permittivity tensor transforms the modes into "quasi-TM" and "quasi-TE" modes. Besides the TM components, the quasi-TM mode also contains TE components. Because all six field components are non-zero for both modes, those modes can be excited by the incident light of any polarisation.

The mode wavenumber of the quasi-TM and quasi-TE modes in the longitudinally magnetised investigated structure is proportional to $g^2$. As shown in Section $2.3$, for typical values of $g$ the Fano resonance shift induced by the applied magnetic field is small and therefore results only in small modifications of the transmittance/reflectance spectra \cite{NKos13}. However, a considerably larger effect is expected to originate from the magnetisation-induced changes of the field distributions of the modes. In this case, the key point is the appearance of the TM components in the quasi-TE mode, which makes it possible to excite this mode by light of TM polarisation. The excited quasi-TE mode takes a fraction of the incident optical energy, changing the overall absorbed energy by a factor proportional to $g^2$.

For zeroth order of diffraction, the considered magneto-optical intensity effect can be described by the relative difference between the transmittance coefficients $T_{\rm{M}}$ and $T_{\rm{0}}$ of the magnetised and the demagnetised structure: $\delta = (T_{\rm{M}}-T_{\rm{0}})/T_{\rm{0}}$ observed at the TE mode frequencies.  Because the longitudinal magnetisation of the investigated structure also modifies the field of the TM modes by inducing TE components, one should expect intensity variation at the TM-mode resonances as well.

This effect is called the "longitudinal magneto-photonic intensity effect" (LMPIE). The LMPIE should be viewed as a novel effect. For example, a similarly defined intensity-related effect, but of different origin, was studied in conventional magneto-optical ferromagnetic films, and was called orientational effect (see, e.g., Ref. \cite{Kri69}). However, for illumination polarised along the magnetisation, the orientational effect vanishes. 

In Ref. \cite{Bel13_LMPIE}, the LMPIE was observed in transmission. Intensities of the zeroth-diffraction-order transmitted light for demagnetised and longitudinally magnetised MPCs were compared to determine the parameter $\delta$ that characterises the LMPIE. Firstly, it was verified that no intensity modulation occurs for the bare magnetic film [Fig. $4$(a), green curve]. Secondly, the LMPIE for the nanostructure illuminated with normally incident TM-polarised light was measured and it was found that the longitudinally applied magnetic field resonantly increases the transparency of the nanostructure by $24\%$ at $840$ nm [Fig. $4$(a)]. Numerical modelling results [blue curve in Fig. $4$(a)] reproduced the experimental data with good accuracy.

\begin{figure}[htbp]
\centering\includegraphics[width=12cm]{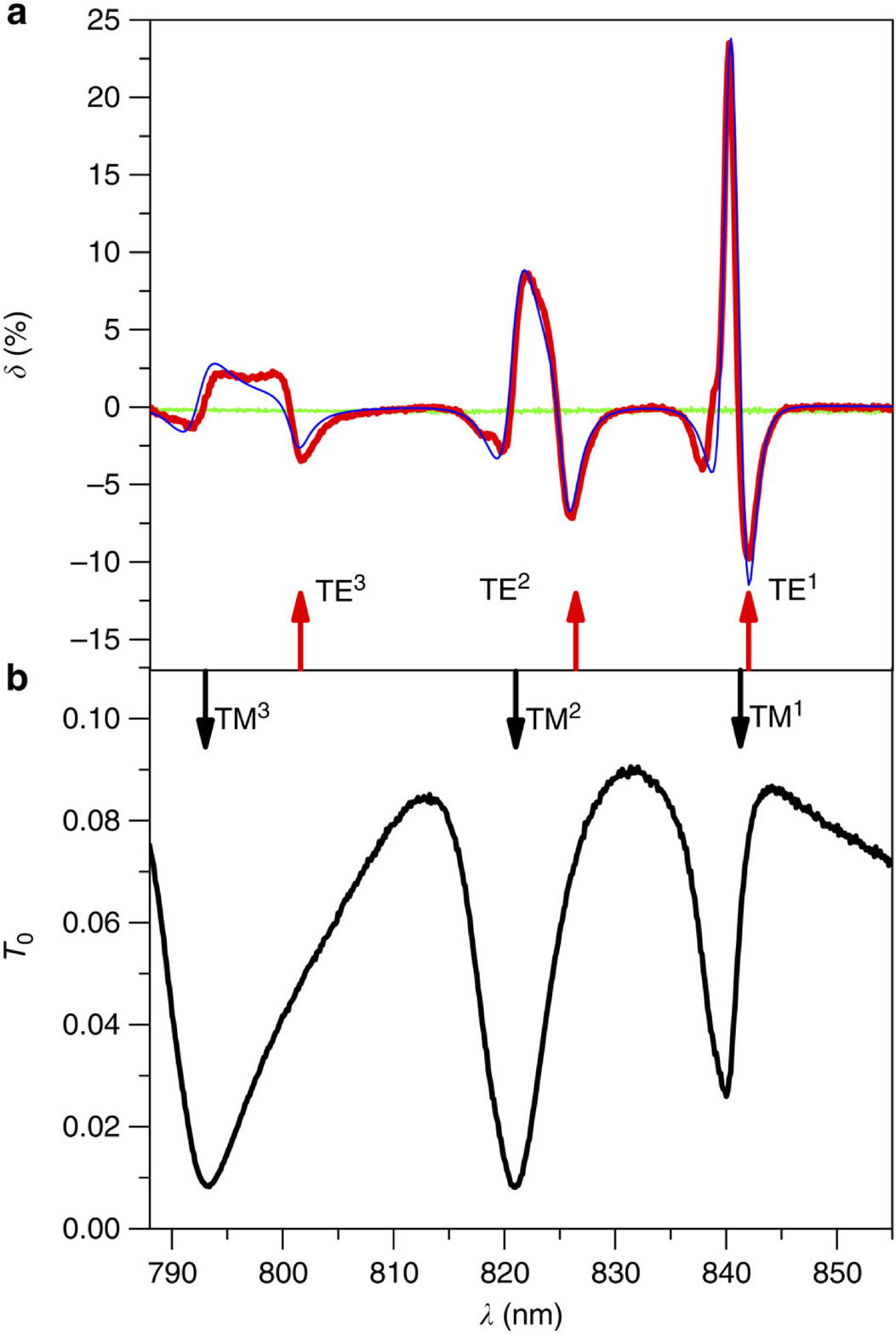}
\caption{(a) Spectrum of the LMPIE when a static magnetic field reaching almost the saturation value. Blue curve shows calculated $\delta$. There is no LMPIE for the bare magnetic film (green curve). (b) Spectrum of the optical transmittance for the demagnetised structure. Black and red arrows indicate calculated spectral positions of the quasi-TM and quasi-TE resonances, respectively. The modes are denoted by the number of their $H_{\rm{y}}$ or $E_{\rm{y}}$ field maxima along the $z$ axis. The light is TM-polarized and hits the sample under normal incidence. Reproduced with permission from Ref. \cite{Bel13_LMPIE}.}
\end{figure}


\section{Prospectives of Magneto-Plasmonics in Spin-Wave and Magnonic Applications}

Let us now discuss how the plasmon-enhanced magneto-optical effects can be used in the emerging areas of spin wave technology and magnonics. Let us start with a simple example. 

A personal computer (PC) consumes $60-250$ W of power. There are more than one billion PCs in use around the world. The total power consumed and dissipated by PCs is alarming and they leave a considerable carbon footprint \cite{Kur08,Sam13}. Therefore, the development of energy-efficient computing devices is recognised as a priority by Intel, Microsoft, and Google.

Significant reduction of power consumption in PCs can be achieved by using magnetic nanodevices - nonvolatile nanomagnetic logics \cite{Sta14}, spin-wave logics \cite{Her04,Kos05,Her07,Sch08,Khi10,Din12}, magnetoresistive random-access memory (MRAM) and spin-transfer torque magnetic random-access memory (STT-MRAM) \cite{Sta14}, which store information magnetically and manipulate it using magnetisation dynamics. The fundamental motion of the vector of magnetisation is its precession about the equilibrium direction in the material [Fig. $5$(a)]. Spin waves [Fig. $5$(b)] represent collective magnetisation precession, which is non-uniform in space \cite{spin_wave_confinement,Mak15}. In spin waves, the amplitude of dynamic magnetisation is small. This property is used in spin wave microwave devices and logics. Magnetic memory and logics based on the reversal of the magnetisation vector [Fig. $1$(c)] employ large-amplitude precessional dynamics to switch the direction of magnetisation.

\begin{figure}[htbp]
\centering\includegraphics[width=16cm]{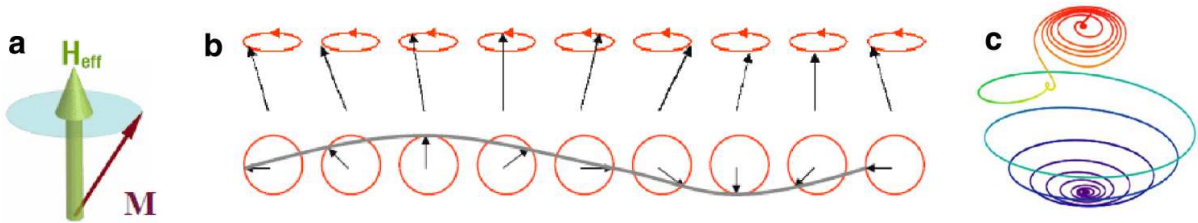}
\caption{ (a) Dynamic precession of the magnetisation vector \textbf{M}, (b) spin wave (one complete precession distributed over a chain of spins), and (c) trajectory of \textbf{M} in the magnetisation reversal process. Reproduced with permission from Ref. \cite{Wan09}.}
\end{figure}

High-quality nanopatterned ferromagnetic media - magnonic crystals \cite{Kru10,Ser10,Gub10,Len11,Sta14,Kra14} provide full control of dynamic magnetisation in a fashion similar to the control of light using photonic crystals \cite{Weh08}. Magnonic crystals are used as the "building blocks" of magnetic RAM and spin wave logic schemes integrated with spin transfer torque oscillators and switches \cite{Sta14,Hir14}. These technologies are promoted actively by semiconductor giants \cite{semiconductor_roadmap}.

The interaction of light with dynamic magnetisation is of utmost importance for the development of novel magnetic nanotechnologies. For example, the main challenge in the development of magnetic memory and logics is the real-time read-out of the logic state. The most probable technological solution is the use of magneto-optical Kerr effect (MOKE) spectroscopy, which allows probing dynamic magnetisation in real time \cite{Kir10,Len11}. MOKE spectroscopy manifests as a magnetisation-induced change in the intensity or polarisation of incident light upon reflection from the magnetised medium \cite{Zvezdin_Kotov}. This effect has been exploited in commercial magneto-optical devices such as magneto-optical discs \cite{Zvezdin_Kotov}.

\begin{figure}[htbp]
\centering\includegraphics[width=14cm]{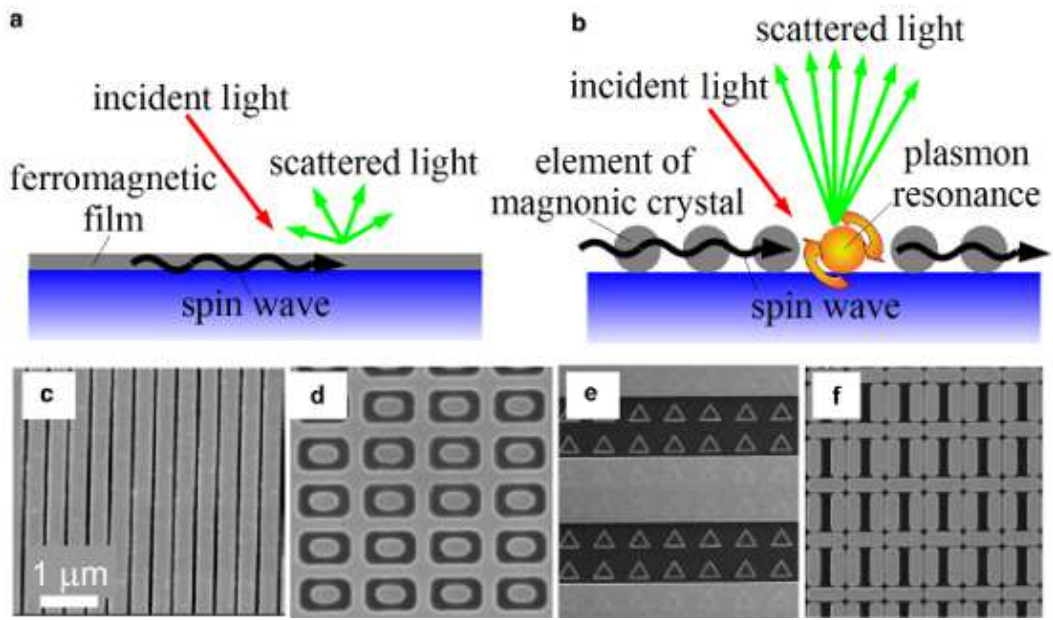}
\caption{ (a) Light interaction with a spin wave in a uniform ferromagnetic thin film sitting on top of a dielectric substrate. (b) Excitation of the plasmon resonance in a ferromagnetic magnonic crystal leads to enhancement of the intensity of light scattered from the spin wave. (c-f) Some examples of plasmon resonance-supporting Permalloy magnonic crystals fabricated at the National University of Singapore (see, e.g., Ref. \cite{ding_thesis}). Reproduced with permission from Refs. \cite{NKos13,triangles,anti_rings,dolmen}}
\end{figure}

However, as shown schematically in Fig. $6$(a), the natural interaction between light and dynamic magnetisation is weak. (For a more detailed discussion see, e.g., Ref. \cite{Arm13}). The strength of the interaction between light and dynamic magnetisation can be increased by exploiting magneto-plasmonic effects, which is possible because plasmon-enhanced magneto-optical response has been achieved in the same nanostructures which are used as magnonic crystals. The plasmon-enhanced light scattering from spin waves is schematically shown in Figs. $6$(a-b). One sees that the excitation of surface plasmons is not possible in the case of a continuous Permalloy ferromagnetic thin film. Consequently, there is no plasmon resonance and the enhancement of the light-magnetic-matter interaction is impossible.

Of course, one can cover the Permalloy film by a prism (see Fig. $10$ in Ref. \cite{Arm13}) in order to excite plasmons under the Attenuated Total Reflection (ATR) geometry \cite{Rae88}. Alternatively, in order to excite plasmons one can nanopattern the film as shown in Figs. $6$(c-f). The structures shown in these figures are actually Permalloy magnonic crystals fabricated by deep UV lithography followed by lift-off \cite{Ade08,ding_thesis}. We already demonstrated that a magnonic crystal [Fig. $6$(a)] combines good optical and magneto-plasmonic properties (Section $2.3$). The magnonic crystals shown in Figs. $6$(d-f) also support plasmon resonances.

For the sake of illustration, let us consider a dolmen-like structure [Fig. $6$(f)] \cite{dolmen}. It is noteworthy that dolmen-like structures play important role in modern plasmonics thanks to their unique optical properties (see, e.g., Ref. \cite{Ver09}). To date, magnonics experiments have been conducted on relatively large arrays of identical dolmens, and the dimensions of these dolmens are comparable with the wavelength of incident light \cite{dolmen}. However, from the viewpoint of magnetism, the dimensions of the dolmens can be reduced readily to an optical subwavelength, which is the case of the dolmens in Ref. \cite{Ver09}.

Figure $7$ shows the simulated absolute values of the electric field in one unit cell of a periodic array of gold and Permalloy dolmens with the same dimensions as in Ref. \cite{dolmen}. The wavelength of normally incident light is $632.8$ nm. One can see the excitation of plasmon modes in the $50$ nm-wide gaps between the individual elements of the dolmens. Naturally, light absorption in Permalloy is larger than in gold. Consequently, amplitude of the electric field in the gaps of the Permalloy structure is smaller than that in the gold structure. Moreover, one can see that in the Permalloy structure the propagation length of plasmons localised in the gaps is significantly smaller. Nevertheless, simulations suggest that in dolmen-like magnonic crystals (and also in other types of magnonic crystals) plasmons may contribute to the magneto-optical response. Consequently, their contribution should be taken into account in order to facilitate understanding of the magnetisation reversal processes in magnetic nanostructures conducted by means of different modalities of the magneto-optic Kerr effect magnetometry and microscopy \cite{San96,Vav00,Gub05,Lisa}.

\begin{figure}[htbp]
\centering\includegraphics[width=12cm]{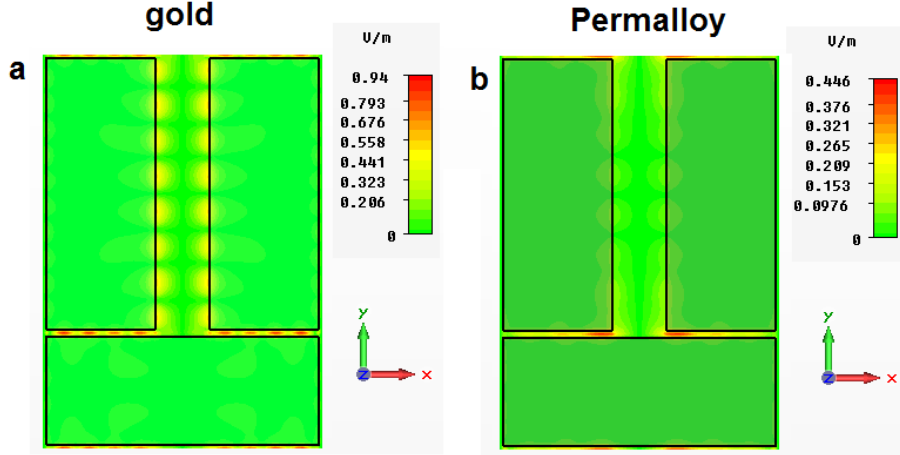}
\caption{ Simulated absolute values of the electric field in one unit cell of a periodic array of (a) gold and (b) Permalloy dolmens with the same dimensions as in Ref. \cite{dolmen}. The wavelength of normally incident light is $632.8$ nm. The width of the gaps between the individual elements of the dolmens is $50$ nm. Simulations were conducted by using commercial CST Microwave Studio software.}
\end{figure}

Apart from the MOKE magnetometry and spectroscopy, one also needs a method with which to probe magnetisation dynamics simultaneously in the frequency and wave-vector domain as well as to map directly the intensity of magnetisation at the nanoscale. Brillouin light scattering (BLS) spectroscopy is the most suitable method \cite{Dem94,Hillebrands1,Hillebrands2,Gub10,Len11}. BLS spectroscopy has been described as the reflection of incident light from a moving Bragg-diffraction grating produced in a magnetic medium due to propagating spin waves \cite{Bor82,Dem94}. As a result, a portion of the scattered light is shifted by the frequency of the spin waves, i.e., it is scattered in an inelastic manner.

In a typical BLS experiment (see, e.g., Ref. \cite{Gub10}), the dispersion relationship of spin waves is measured with \textit{p}-polarised monochromatic light incident at angle $\theta$ [Fig. $8$(a)]. (The angle $\theta$ is controlled in a broad range because it is linked to the wave vector $k_{\rm{SW}}$ of the probed spin wave by the relationship $k_{\rm{SW}} = (4\pi/\lambda)sin\theta$, where $\lambda$ is the wavelength of incident light.) Importantly, the same polarisation of incident light is required for the excitation of surface plasmons. Consequently, by analogy with the plasmon-enhanced TMOKE response, the conditions for the resonant enhancement of the BLS signal due to surface plasmons can be met in BLS measurements of metallic ferromagnetic magnonic crystals. 

A typical BLS spectrum shows the measured intensity of the scattered light against the measured frequency shift [Fig. $8$(b)]. The elastically scattered incident light is in the centre of the spectrum. The Stokes and anti-Stokes peaks correspond to inelastically scattered light. The frequency shift corresponding to the maxima of these peaks is denoted by dots in the dispersion diagram shown in Fig. $8$(c). In this figure, the lines are the theoretical prediction and the shaded areas are the magnonic band gaps.
	
Technically, the central elastically scattered peak in the BLS spectra in Fig. $8$(b) is the reflectivity of light $R$ detected in measurements of static transverse MOKE. Static MOKE magnetometry is very useful for probing the state of magnetic logic elements. The logic state changes as a result of magnetisation reversal, and can be "0" or "1", which correspond to opposite directions for the magnetisation vector. Time-domain MOKE spectroscopy allows probing ultrafast (sub-nanosecond scale) magnetisation dynamics with high temporal resolution. This makes time-domain MOKE spectroscopy very useful for investigating ultrafast magnetic switches and multiplexors driven by time-domain signals of different shape and duration \cite{Kir10,Hillebrands1}.

\begin{figure}[htbp]
\centering\includegraphics[width=16cm]{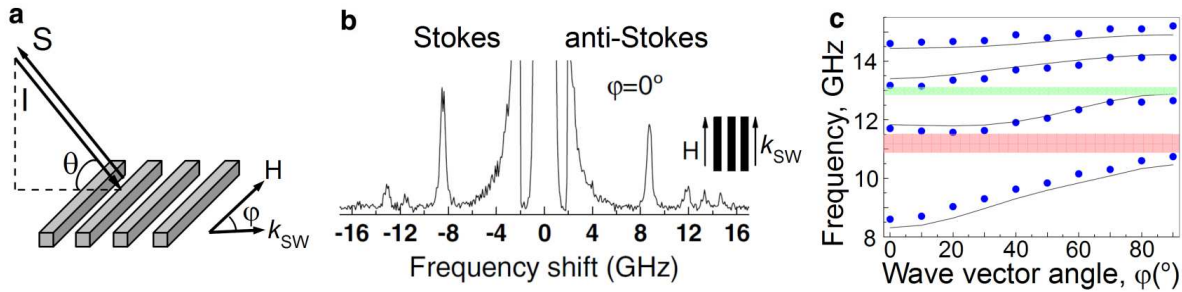}
\caption{ (a) Schematic of a typical BLS experiment. (b) Example of a measured BLS spectrum. (c) Experimental (dots) and theoretical (lines) dispersion curves of the Permalloy magnonic crystal shown in Fig. $7$(a). Reproduced with permission from Ref. \cite{Gub10}.}
\end{figure}

It is noteworthy that several attempts have been made to increase the BLS from acoustic waves using plasmons \cite{Fuk80,Lee90,Ute07,Joh09}. Moreover, the dynamics of acoustic phonons generated by femtosecond impulsive optical excitation can be clearly resolved by a surface plasmon technique, with enhanced sensitivity orders of magnitude higher than regular optical probe measurements \cite{Ext88,Wan07,Tem12,Khokhlov}.  However, one needs to keep in mind that whereas the BLS from acoustic waves is photon-phonon scattering, the BLS from spin waves is photon-magnon scattering. Most significantly, to excite plasmons one needs to fabricate large-area metallic nanostructures on top of the acoustic wave waveguide. In contrast, magnonic crystals inherently support plasmons without the need for additional metallic nanostructures \cite{NKos13}.

Nevertheless, in the remainder of this Section we discuss the results from Ref. \cite{Ute07} that demonstrates the surface-plasmon enhancement of BLS from acoustic waves in gold-discs arrays fabricated on top of glass [Fig. $9$(a)]. Hereafter, the reader will see that this result is encouraging for further research on the plasmon-enhanced BLS from spin waves.

In Ref. \cite{Ute07}, BLS anti-Stokes spectra were taken at $60^{\rm{o}}$ incident angle for the gold continuous film on glass, a bare glass substrate with the same composition, and a nanodisc array on glass. It is noteworthy that surface plasmon excitations do not occur when light is incident on a continuous gold film at the gold/air interface because the dispersion curves of light and surface plasmons do not overlap \cite{Rae88}. For the gold film on glass two peaks were detected, which correspond to the well known Rayleigh-like and Sezawa-like modes. Also, additional peaks appeared in the spectrum from the gold nanodisc array. These peaks had no corresponding modes in the gold film. 

Figure $9$(b) shows the measured enhancement factors $\Gamma$ for the Rayleigh-like and Sezawa-like modes as a function of the nanodisc diameter at an angle of $60^{\rm{o}}$. One can see that the maximum enhancement for the Rayleigh-like mode occurs for the smallest nanodisc size, but varies just $\sim10\%$ over the $20$ nm range of diameters. However, measurements demonstrate the opposite trend for the Sezawa-like mode enhancement, for which the largest values of the enhancement factor occurs with the largest nanodiscs. Interestingly, the demonstrated enhancement is consistent with the simulated values of electric-field intensity on the top and the bottom surfaces of the gold nanodiscs. (These values were normalised to the incident intensity.) It suggests that by improving the light-focusing properties of the discs one can improve the enhancement factor $\Gamma$. Most significantly, unlike in the case of the light-magnetic matter interaction, the interaction of light with acoustic waves does not involve magneto-optical effects. As discussed throughout this review, the strength of magneto-optical effects can be considerably enhanced by means of plasmon resonances. Consequently, in magnonic BLS experiments one may expect even larger values of $\Gamma$ for spin wave modes.

\begin{figure}[htbp]
\centering\includegraphics[width=8.5cm]{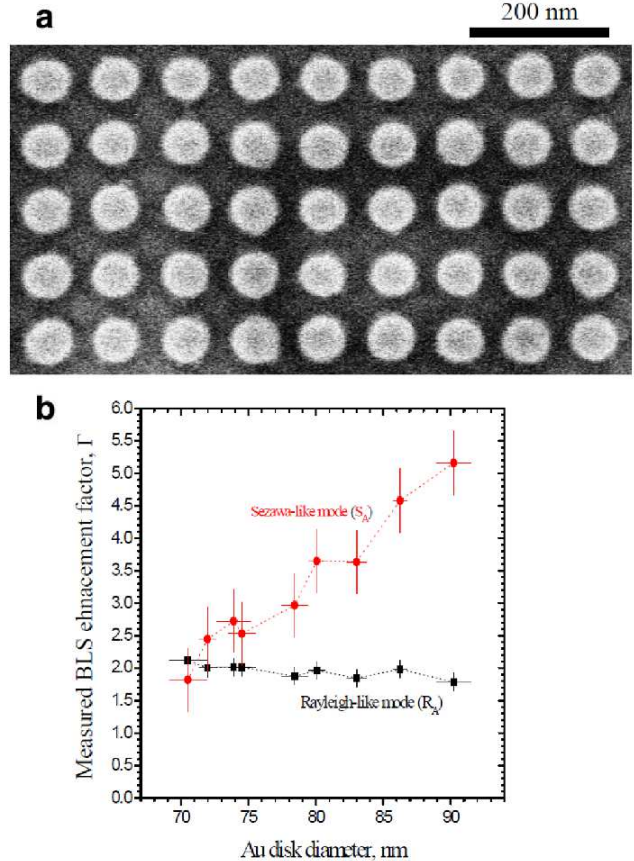}
\caption{ (a) Image of a section of a gold nanodisc array on glass. (b) Acoustic BLS enhancement factor $\Gamma$ the Rayleigh-like and Sezawa-like modes as a function of the nanodisc diameter. Reproduced with permission from Ref. \cite{Ute07}.}
\end{figure}


\section{Enhanced Magneto-Optical effects in All-Magneto-Insulating Nanostructures}

Since all metals absorb light in the visible and infrared spectral ranges, maximum efficiency of metallic plasmonic devices can be achieved only by using metals with the lowest absorption cross-section. Therefore, most of the plasmonic devices are made of gold or silver because these two metals exhibit the lowest absorption losses at optical frequencies. (However, losses remain considerable even if gold or silver are used as the model material.) For example, transmission spectra of a gold grating with exactly the same dimensions as the Permalloy grating in Fig. $3$(a) were investigated in Ref. \cite{NKos13}. (Of course, the gold grating does not exhibit magneto-optical properties; it was used as a reference to better understand the optical response of the Permalloy grating.) As shown in Fig. $2$ in Ref. \cite{NKos13}, the reflectivity of the gold grating is more than $3$ times larger than that of the Permalloy grating. A combination of such a high reflectivity with a strong plasmon-enhanced TMOKE response would be advantageous for practical applications.

It is noteworthy that inevitable absorption losses in metals at optical wavelengths has already motivated researchers in the photonics sector to investigate diffraction gratings (see, e.g., Ref. \cite{Wan93}) as well as metamaterials and nanoantennas based on alternative materials such as semiconductors \cite{Nai13} and dielectrics \cite{Kuz12,Kra13}. Naturally, the same strategy of substituting metals with dielectrics works in the field of magneto-optics. For example, Bai et. al. \cite{Bai06} demonstrated theoretically strong polarisation conversion effects, including large Kerr rotation and ellipticity, in zero-order reflection of light from a BIG grating embedded in a dielectric material. The field localisation that gives rise to the enhanced magneto-optical effects was achieved due to the excitation of leaky guided waves, which propagate in (and around) the high-index BIG layer in directions perpendicular to the external static magnetic field. It was shown that whereas the resonant interaction with the incident field results in narrow-band reflection spectra, large magneto-optical polarisation conversion effects result from the splitting of the degeneracy of right- and left-circularly polarised eigenmode resonances with highly dispersive phase spectra near the resonance peak.

Similar results were obtained by Marinchio et. al. \cite{Mar14} who demonstrated theoretically that subwavelength nanowire gratings made of arbitrary anisotropic materials can present a resonant response at specific geometric conditions. They showed that for a given polarisation of incident light a resonant magneto-optical response can be obtained by tuning the incidence angle and grating parameters to operate near the resonance condition for the opposite polarisation. These results are important for the understanding and optimisation of magneto-optical structures and devices based on resonant subwavelength gratings, which is used in sensing applications \cite{Qin15}.

We would like to note that the gratings investigated in Refs. \cite{Bai06,Mar14} are subwavelength, i.e. their period is significantly smaller than the wavelength of incident light. In such gratings, at the resonance wavelength the zero-order reflected and transmitted light is present in the far field, with higher-order diffraction modes suppressed (see Fig. $10$). The high reflectivity exhibited by subwavelength gratings depends on the angle of incidence and it is attributed to the excitation of a waveguide mode through phase matching by the grating. The waveguide mode then recouples back (again through grating phase matching) to radiative modes and produces the observed high reflection \cite{Wan93}. The recoupling occurs at the frequencies of the resonant Wood's anomaly \cite{Hes65}, which in turn is linked to the Rayleigh wavelength at the given angle of incidence. Furthermore, the lineshape of the resonance peaks in the reflectivity spectrum of the grating is asymmetric due to Fano interference \cite{Hes65,Mir10}.

The complex resonance behaviour of the proposed subwavelength dielectric grating is a mechanism that differs it from conventional magneto-optical diffraction gratings having a large period as compared with the wavelength of incident light (see, e.g., Ref. \cite{Sou98}). Moreover, conventional gratings were made of magneto-insulating wires on a silicon substrate. The application of these materials for the fabrication of nanostructures was a technological challenge until the recent progress in nanostructuring of BIG and YIG (see, e.g., Ref. \cite{Ame14}) and similar materials. New fabrication technologies enable the formation of complex (two- and three-dimensional) optical and magnetic nanostructures in YIG, which opens up opportunities to development novel devices for optical communication and photonic integration.

Consequently, in the theoretical work Ref. \cite{Mak_14OE} it was proposed to employ BIG or YIG as one of the constituent materials of a subwavelength grating having exactly the same dimensions as the all-Permalloy grating in Fig. $3$(a). Also, in order to overcome the drawback of the all-Permalloy grating - a low reflectivity of just $\sim20\%$ -  it was suggested to use silicon as the second constituent material to fill the gaps between the BIG (or YIG) nanowires. It was shown by means of rigorous simulations that, for the same angle of incidence as in Ref. \cite{NKos13}, one can achieve a one order of magnitude enhancement of the TMOKE response as compared with a continuous magnetic film. Furthermore, as shown in Fig. $1$ of Ref. \cite{Mak_14OE}, the silicon nanowires play the crucial role in achieving a high reflectivity of $\sim80\%$. (Recall that for the all-Permalloy grating the reflectivity is just $\sim20\%$.) Most significantly, as shown in Fig. $10$, both the TMOKE response of the all-magneto-dielectric grating and its reflectivity increase as the angle of incidence is decreased. This increase is not readily attainable using magneto-plasmonic gratings made of magneto-insulating films and gold nanowires.

To conclude this Section we would like to note that unique advantages of non-magnetic dielectric photonic nanostructures over non-magnetic metallic plasmonic structures are their low absorption losses. These advantages can boost performance of many photonic devices in the visible and near-IR frequency spectra \cite{Kuz12,Kra13}. By analogy, the application of all-magneto-dielectric nanostructures instead of metallic ones can is expected to boost performance of many magneto-optical devices.

However, the application of similar ideas in the field of magnonics is not straightforward. As shown in Section $2.3$, Permalloy and other magnetic metal alloys are very important for magnonics and spintronics. Moreover, despite the recent progress \cite{Ame14}, the fabrication of high-quality YIG nanostructures remains a challenge and also YIG is relatively expensive. On the other hand, as discussed in Ref. \cite{Ser10}, until certain technical obstacles are overcome in modern magnetic alloy materials, YIG remains a valuable source of insight and the interest in YIG-based magnonic systems is particularly strong. Consequently, we anticipate the appearance of hybrid structures combining magnetic and non-magnetic conductive and insulating materials. Indeed, such structures have been proposed in the recent works Refs. \cite{Qin15,Kho15}.

\begin{figure}[htbp]
\centering\includegraphics[width=12cm]{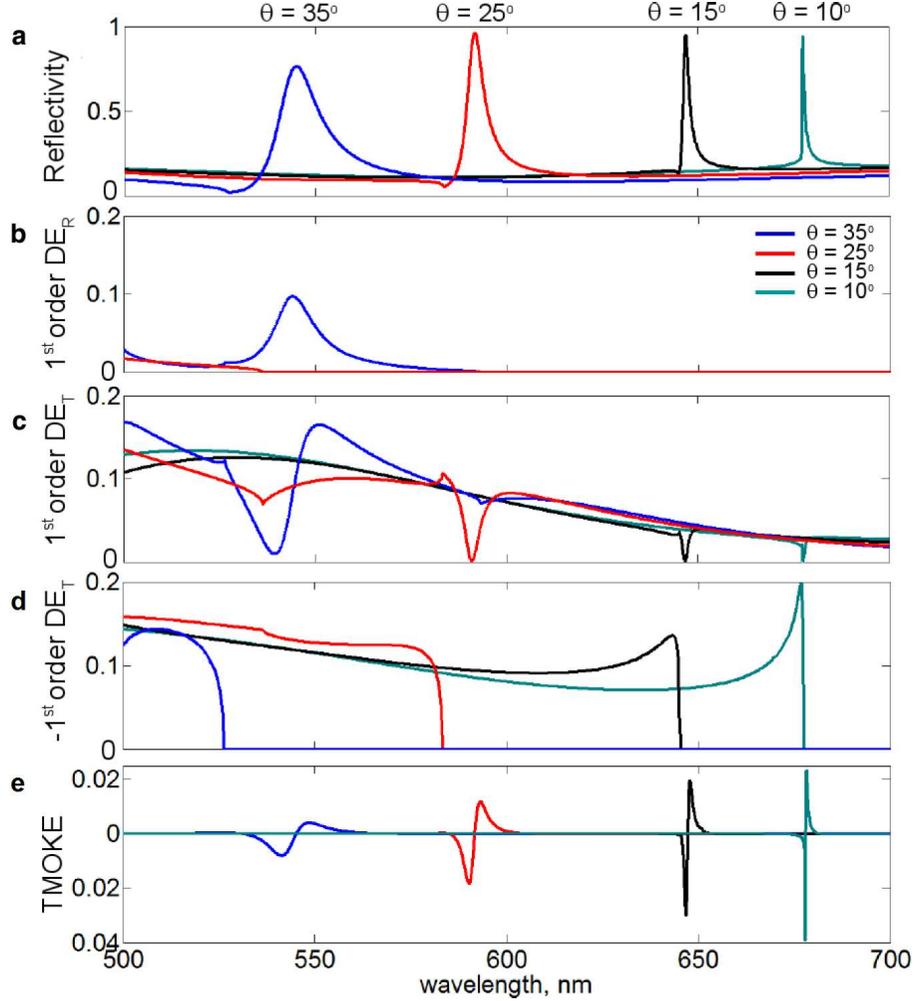}
\caption{ (a) Reflectivity and (b-d) reflected ($DE_{\rm{R}}$) and transmitted ($DE_{\rm{T}}$) diffraction efficiency for the $+1$st and $−1$st orders of the Bi:YIG-Si grating at different angles of incidence $\theta$. (e) TMOKE response of the grating at different $\theta$. The minor secondary peaks in the spectra occur at the Rayleigh wavelengths. Reproduced with permission from Ref. \cite{Mak_14OE}.}
\end{figure}


\section{Plasmon-Enhanced Inverse Magneto-Optical Effects and All-Optical Magnetisation Switching}

\subsection{Plasmon-Enhanced Inverse Faraday Effect}

Ultrafast all-optical control of a medium magnetisation at the subpicosecond time scale \cite{Kim05,Kir10,Lam14} is of utmost importance for the development of novel magnetic data storage systems and generation of spin waves via light \cite{Sat12}. The optical way to control the magnetisation is enabled by the phenomenon in which a circularly polarised light induces static magnetisation in a gyrotropic medium. This phenomenon is called the inverse Faraday effect, which was predicted theoretically in the 1960s \cite{Pit61,Zie65}. (Note that not only the Faraday effect has its inverse counterpart, but there is also the inverse Cotton-Mouton effect \cite{Riz10}.) This effect is nonlinear \cite{Zvezdin_Kotov,Pop94}, which should be taken into account when spin dynamics is manipulated directly and coherently by using multiple laser pulses \cite{Per06}.

The Faraday effect \cite{Zvezdin_Kotov,Kir10} is observed as a rotation of the polarisation plane of light transmitted through a magnetic medium and it can be expressed as $\alpha_{\rm{F}} = \frac{\chi}{n}$\textbf{M}$\cdot$\textbf{k}, where $\alpha_{\rm{F}}$ is the specific Faraday rotation, \textbf{M} is the magnetisation, $n$ is the refractive index, \textbf{k} is the wave vector of light, and $\chi$ is the magneto-optical susceptibility. The inverse Faraday effect, which is observed as the induction of a static magnetisation \textbf{M}$_{0}$ by a high-intensity laser radiation, is determined by the same magneto-optical susceptibility $\chi$ as \textbf{M}$_{0} = \frac{\chi}{16\pi} [$\textbf{E}$(\omega) \times$ \textbf{E}$^{*}(\omega) ]$, where \textbf{E}$(\omega)$ and \textbf{E}$^{*}(\omega)$ are the electric field of the light wave and its complex conjugate, respectively \cite{Kim05,Her05,Kir10,Khokhlov}. 

Importantly, in the case of the inverse Faraday effect $\chi$ is a ratio between the induced magnetisation and the laser intensity. Consequently, as discussed in Refs. \cite{Kim05,Kir10}, optical control of magnetisation is the most efficient in materials with high values of the Faraday rotation per unit magnetisation. Furthermore, $\chi$ is allowed in all media regardless their crystallographic and magnetic structures. Most significantly, the inverse Faraday effect does not require light absorption. This implies that the effect of light on magnetisation is non-thermal and can be considered as instantaneous because it takes place on a femtosecond time scale. The first experimental demonstration of such non-thermal ultrafast optical control of magnetisation was done by Kimel et. al. \cite{Kim05}, who used the inverse Faraday effect to excite magnetisation dynamics in $60$ $\mu$m thick DyFeO$_{3}$ magnetic samples, in the hundreds of GHz frequency range. They demonstrated that when a circularly polarised light pulse of high intensity (pump pulse) is focused on a magnetic material, a spin precessional motion starts within the light spot via the inverse Faraday effect, whereby an effective magnetic field is generated along the beam direction. 

Building up on the results in Ref. \cite{Kim05}, Satoh et. al. \cite{Sat12} investigated YIG samples. (Recall that YIG is often employed in magnonic and spintronic devices because of intrinsically low magnetic damping in YIG \cite{Ser10}). Initially, the magnetisation laid in the plane parallel to the sample surface.  When a circularly polarised pump beam with a pulse width of $\sim100$ fs was focused into a circular spot on the sample surface, spin precessional motion occurred within the spot via the inverse Faraday effect. The magnetisation deviated from the plane and the out-of-plane component became nonzero. This out-of-plane magnetisation component was detected by the Faraday polarisation rotation of a low-intensity linearly polarised pulse (probe pulse). A time-resolved pump-probe experiment was carried out by measuring the Faraday rotation angle of the probe pulse, which is time-delayed with respect to the pump pulse. The precessional motion propagated in two dimensions out of the pump light spot in the form of a spin wave [Fig. $11$(a)]. By scanning the relative position of the probe light spot on the sample with respect to the pump light spot, it was possible to observed time- and space-resolved spin wave propagation [Fig. $11$(b)]. It was found that the wavelength of the spin wave was $200-300$ $\mu$m, and the group velocity was about $100$ km/s \cite{Sat12_SPIE}. Numerical simulations reproduced the experimental result with good accuracy [Fig. $11$(c)].

\begin{figure}[htbp]
\centering\includegraphics[width=16cm]{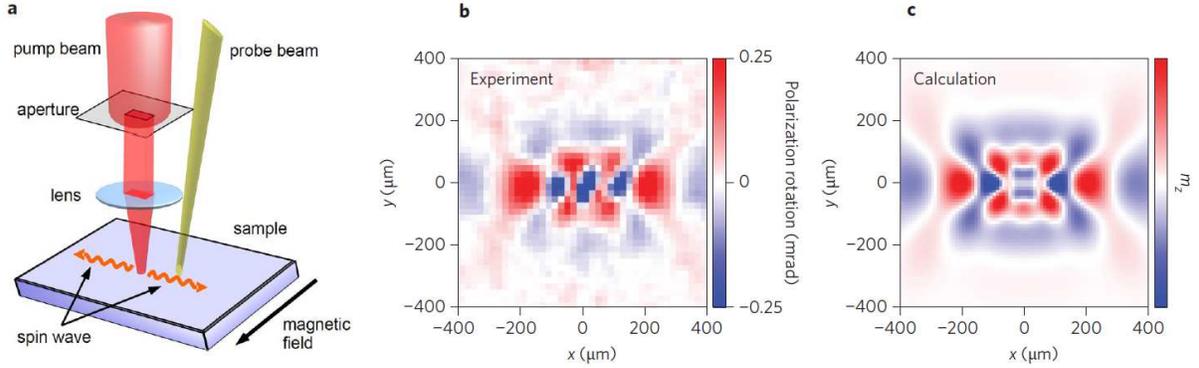}
\caption{ (a) Experimental configuration for directional control of spin wave emission. (b) Two-dimensional spin waveforms obtained by (b) experiment and (c) simulation. Reproduced with permission from Refs. \cite{Sat12, Sat12_SPIE}.}
\end{figure}

As discussed above, in the experiment from Ref. \cite{Sat12} spin precessional motion occurred within the spot via the inverse Faraday effect. Therefore, one may hypothesise that, similar to Ref. \cite{Sti10} in which magnetic recording using an integrated plasmonic antenna was demonstrated, the efficiency of all-optical spin wave excitation may be improved by enhancing the inverse Faraday effect by means of plasmons. This can be done, for example, by fabricating a plasmonic nanoantenna (for a review see, e.g., Ref. \cite{Kra13}) on top of the magnetic sample.

However, as discussed in Ref. \cite{Guy15}, one should keep in mind that the coupling of a laser pulse to nanoantennas with a small footprint is challenging. Hence, one needs to employ arrays of nanoantennas or single nanoantennas that have a large footprint comparable with the spot of the laser beam. In both cases, one of the most probable candidates are the so-called tapered Yagi-Uda-type arrayed nanoantennas, which combine a large footprint with high efficiency in a broad spectral range \cite{Mak12,Sta12}. Finally, we would like to note that arrayed nanoantennas can be made by using dielectric nanoparticles (for a review see, e.g., Ref. \cite{Kra13}). The potential advantage of such all-dielectric nanoantennas is their high coupling efficiency achievable due to low absorption losses in the constituent dielectric material.   

\subsection{Plasmon-Enhanced Inverse Transverse Magneto-Optical Effect}

As an alternative strategy for the enhancement of light coupling to the magnetic medium one can employ all-ferromagnetic nanostructures (i.e., magnonics crystal) and combine them with nanostructures made of plasmonic metals (i.e., gold or silver). This combination was investigated by Belotelov and Zvezdin \cite{Bel_iTMOKE}, who revealed the existence of the inverse TMOKE in continuous magnetic films (Fig. $12$) and showed that the strength of this effect can be increased by exploiting plasmonic resonances in magnetic nanostructures (Fig. $13$).   

It is shown in Section $5.1$ that the cross product $[$\textbf{E}$(\omega) \times$ \textbf{E}$^{*}(\omega)]$ is non-zero for elliptically polarised light. In contrast, in the TMOKE configuration \cite{Arm13} the incident light is linearly polarised. However, as discussed in Ref. \cite{Bel_iTMOKE}, due to the boundary conditions between a dielectric medium and a medium with absorption or with negative permittivity linearly polarised light is converted into  elliptically polarised light. This implies that the cross product $[ \textbf{E}(\omega) \times \textbf{E}^{*}(\omega) ]$ is non-zero. It is noteworthy that this picture is for the \textit{p}-polarised incident light (see Fig. $12$) and it does not hold for the case of the \textit{s}-polarised light \cite{Bel_iTMOKE}. We also note that, by analogy with the inverse Faraday effect, the inverse TMOKE is also a nonlinear phenomenon. 

\begin{figure}[htbp]
\centering\includegraphics[width=8.5cm]{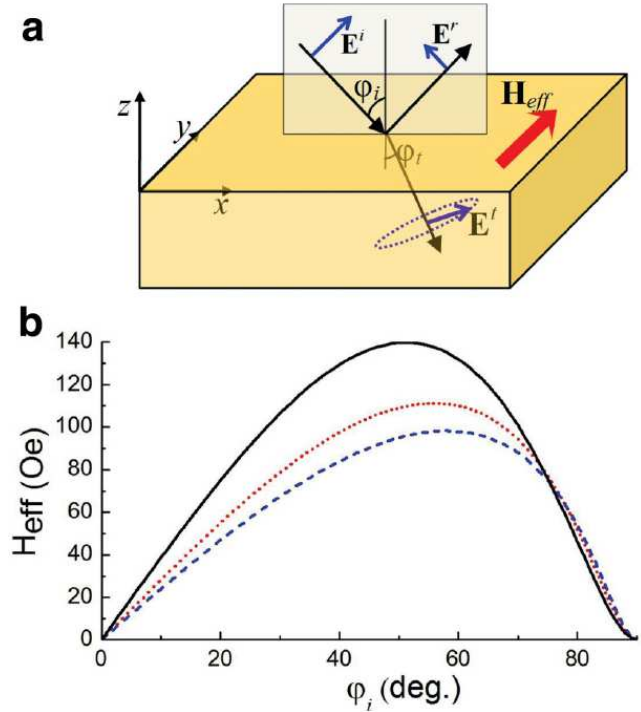}
\caption{(a) Configuration for the inverse TMOKE. \textit{p}-polarised light is obliquely incident on the ferromagnet film and
generates an effective magnetic field $H_{\rm{eff}}$ inside the film. (b) The inverse TMOKE near the surface (at a point $5$ nm in depth) of the magnetic films of iron (black solid line) at $\lambda = 630$ nm, nickel (red dotted line) at $\lambda = 630$ nm, and bismuth iron garnet (BIG) (blue dashed line) at $\lambda = 400$ nm as a function of angle of incidence of light. It is assumed that the samples are illuminated with a $40$-fs duration laser pulse with a peak intensity of $500$ W/$\mu$m$^2$. Reproduced with permission from Ref. \cite{Bel_iTMOKE}.}
\end{figure}

A necessary condition for the appearance of an effective magnetic field in a ferromagnetic medium due to the inverse TMOKE is a decay of the electromagnetic field inside this medium due to optical absorption losses and/or a negative real part of the dielectric permittivity of plasmonic metals. However, for thin ferromagnetic films in which light can be reflected from the bottom film surface, the effective magnetic field does not vanish even for media with a purely real refractive index.

Consequently, one can see an important difference between the inverse TMOKE and the inverse Faraday effect. In the inverse Faraday effect, the effective magnetic field is induced by circularly polarised light and is detected along the wave vector. In contrast, the effective magnetic field induced due to the inverse TMOKE is orientated along the cross product of the incident wave vector and the normal to the surface of the ferromagnetic film. Due to this difference, the application of the inverse TMOKE in optically induced femtosecond magnetism is potentially more promising.

Before we discuss the enhancement of the inverse TMOKE in ferromagnetic nanostructures, we would like to note that, as shown in \cite{Bel_iTMOKE}, the inverse TMOKE is not related to the magnetic dichroism. It is also different from the inverse Cotton-Mouton effect \cite{Riz10}.

The simplest way to excite surface plasmons and enhance the strength of the inverse TMOKE is to use a continuous nickel film and a prism. This configuration were used, e.g., in Ref. \cite{Arm13} to enhance the "direct" TMOKE. (The term "direct" is used to avoid possible confusion with "inverse".) In the case of the inverse TMOKE this approach allowed to increase the induced effective magnetic field by $\sim4$ times \cite{Bel_iTMOKE}. Moreover, a periodic composite of alternating gold and nickel nanostripes makes it possible to increase the effective magnetic field by $\sim6$ times with respect to the plasmon-assisted enhancement in the continuous film. This is achievable because for the incident light wavelength of $630$ nm the permittivities of gold and nickel have close real parts. (But their imaginary parts greatly differ because of significantly larger absorption losses in nickel.) Consequently, the wavelength of the surface plasmons waves at the gold and nickel surfaces are almost the same. (However, the corresponding absorption coefficients are very different.)

Belotelov and Zvezdin also considered the case of an all-nickel one-dimensional grating that is also often called plasmonic crystal. The dimensions of this structure are very similar to those of the one-dimensional Permalloy magnonic crystal in Fig. $3$(a). As discussed in Section $2.3$, microwave magnetic properties of Permalloy are suitable for applications in magnonics, but the application of nickel is  limited. However, magneto-plasmonic properties of the nickel structure are very similar to the properties of the Permalloy structure because Permalloy is a nickel-iron magnetic alloy, with about $80\%$ nickel content. Consequently, the result for the nickel grating presented below is also applicable to a broader class of devices -- magnonic crystals.

The nickel nanostructure concentrates electromagnetic energy density at the surface plasmon resonance conditions, which leads to a local energy enhancement in the near-field region [Fig. $13$(a)]. As a result, the effective magnetic field induced due to the inverse TMOKE is increased and it exceeds $5000$ Oe near the nickel interface. A prominent feature of the inverse TMOKE in the nickel structure is that the effective magnetic field has opposite directions in different parts of the nickel nanowire cross-sections [Fig. $13$(b)], which opens up additional opportunities for the magnetisation control in ferromagnetic nanostructures.  

To conclude this section, we would like to note that similar to the inverse Faraday effect the enhancement of the inverse TMOKE has also been predicted in magneto-insulating structures combined with gold nanowires \cite{Poh13}.

\begin{figure}[htbp]
\centering\includegraphics[width=8.5cm]{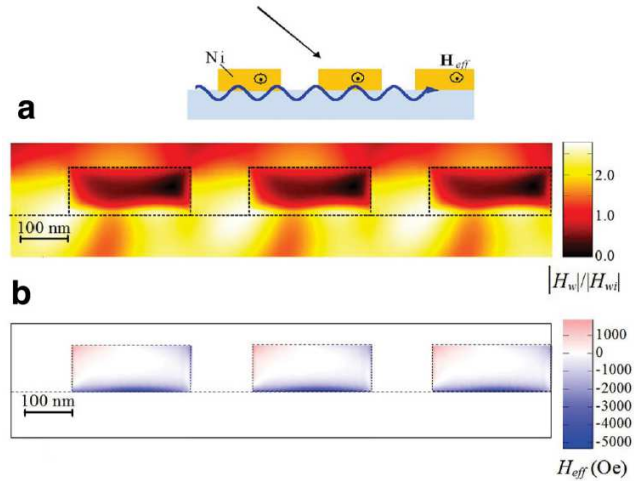}
\caption{(a) Near-field distribution of the absolute value of the electromagnetic wave magnetic field $|H_{\rm{w}}|$ in the plasmonic crystal (shown in the inset above) normalised to the magnetic field of the incident light $|H_{\rm{wi}}|$. (b) Distribution of the effective magnetic field $H_{\rm{eff}}$ generated in the plasmonic crystal by the laser pulse. Three periods of the structure are shown. A laser pulse with a peak intensity of $500$ W/$\mu$m$^2$ is \textit{p}-polarised and is obliquely incident at $10^{\rm{o}}$. Period $d=400$ nm, slit width is $120$ nm, and nickel film thickness is $100$ nm. Reproduced with permission from Ref. \cite{Bel_iTMOKE}.}
\end{figure}


\section{Perspectives of Terahertz Magneto-Plasmonics in Ultrafast Control of Dynamic Magnetisation }

Terahertz (THz) radiation is used in telecommunications, non-invasive imaging, and other areas of science and technology (see, e.g., Ref. \cite{Mit13}). The THz region ($\sim0.1$ THz to $\sim10$ THz) is often described as the last frontier of the electromagnetic spectrum because of the relatively low maturity level of components and systems that operate in this region. As a result, the THz community continues to be active and growing. THz magneto-plasmonics is one of the research directions of this common effort. 

The recent advances in magneto-plasmonics may be split into three groups. The first group includes non-reciprocal surface magneto-plasmon waveguides, gratings, resonators and metamaterials for THz radiation (see, e.g., Refs. \cite{Wan04,Kum13,Con13,Tsi13,Ber13}). This group also includes magnetically tuneable THz isolators \cite{Hu12,Fan12}. The second group is concerned with tuneable magneto-plasmons to achieve efficient control of THz radiation by graphene monolayers. Some papers from this group are cited in the following Section in the context of graphene plasmonics for THz to infrared applications \cite{Jab09,Gri12,Low14,Garcia14}.

The works from the third group are concerned with probing and controlling the magnetisation dynamics in magnetic structures. It is known that in antiferromagnets the spin waves typically occur at $1$ THz \cite{Pim09}. However, in contrast to ultrafast visible-light-based techniques enabling the dynamic magnetisation to be controlled with a temporal resolution of picoseconds to femtoseconds (see above), THz techniques are not yet well developed. At present, the progress in this direction is motivated by numerous potential advantages of using THz radiation instead of light. 

The first advantage is that THz pulses can trigger spin precession on the timescale of subpicoseconds without inducing unwanted electronic excitation or thermal effects \cite{Naka10,Kam11,Yama13}.  For example, the authors of Ref. \cite{Kam11} explored the interplay of the NiO antiferromagnet with highly intense THz pulses, with a time resolution of $8$ fs. A highly intense THz pulse was focused onto the sample (for more details see Fig. $1$ in Ref. \cite{Kam11}). The THz pulse had a broad spectrum from $0.1$ to $3$ THz, fully overlapping with the magnon resonance at $1$ THz. An $8$ fs pulse probed the THz-driven magnetisation dynamics at a variable delay time \textit{t}. In the experiment, the dynamic magnetisation \textbf{M}(\textit{t}) induced by the pulse manifested itself by the Faraday effect, where the projection of \textbf{M}(\textit{t}) onto the propagation direction of the probe pulse caused a transient circular birefringence that rotates the linear probe polarisation by an angle, which value is exclusively caused by the THz pulse \cite{Kam11}. The Fourier transformation of the resulting time trace of the measured Faraday rotation produced a narrow peak at $1$ THz, clearly linking the signal to the high-frequency spin eigenmode in NiO.

An even more interesting effect would be the ultrafast all-optical magnetisation switching in antiferromagnets. This effect was predicted in the theoretical work Ref. \cite{Wie12}. The authors of this work suggested that THz laser pulses can in principle be used to switch the sublattice magnetisation of antiferromagnets on a picosecond scale.

As discussed in Ref. \cite{Con14}, another advantage is that the THz energy range is comparable with the low-energy terms of the magnetic Hamiltonian such as magnetic anisotropy and the Dzyaloshinsky-Moriya interaction. This property of THz radiation provides an opportunity to measure these effects directly. This is important in the development of novel materials for spintronic applications, but often is challenging for other experimental techniques such as intelatic neutron scattering (see Ref. \cite{Con14} and references therein).

Although the possibility to use magneto-plasmons for the improvement of THz-bases control and spectroscopy of spin waves was not the topic in the cited papers, we point out that it should be possible to combine THz plasmonic properties with the capability of THz radiation to control and probe spin waves. Our suggestion is based on recent advanced in an adjacent research direction of THz metamaterials. For instance, Kurihara et. al. \cite{Kur14} demonstrated enhancement of the spin precession of orthoferrite using the magnetic near-field produced by a split-ring resonator (SRR) using the THz spin-pump Faraday probe measurement technique. The precession amplitude was enhanced by $\sim8$ times when the resonance frequency of spin precession was close to the magnetic resonance of SRR. It suggested that the local enhancement of the magnetic near-field in the THz range by metallic structures (see, e.g., Ref. \cite{Mit13}) may be used for further increase in the enhancement factor, leading to the nonlinear control of spin dynamics using THz radiation. 


\section{Relevant Topics}

In this Section we would like to highlight several topics that go beyond the scope of this review paper, but remain  of immediate relevance to magneto-plasmonics and also to a broader research area of non-reciprocal surface waves \cite{Dav13,Dav131,Che14}. For instance, surface plasmons play a very prominent role in the optical response of different nanoscale materials systems that include not only metals and semiconductors but also graphene \cite{Nai13}. It is known that graphene supports surface plasmons at infrared and THz frequencies \cite{Jab09,Gri12,Low14,Garcia14}. Hence, magneto-plasmonic interaction has been predicted theoretically and demonstrated experimentally in graphene structures (see, e.g., Refs. \cite{Loz12,Cra12,Fer12,Yan12,Zhou13,Cha14,Hu14,Had14}).

Plasmon-enhanced magneto-optical effects are also important for the development of Spatial Light Modulators capable of enhancing two-dimensional ($2$D) projection systems and $3$D holographic displays for smartphones, projectors and $3$D television \cite{Aos10}. We would like to discuss this important topic in more detail.

In the stereoscopic-based $3$D displays, viewers can only artificially perceive $3$D and they do not see actual $3$D images, which may result in discomfort or even illness caused by eye fatigue. Holographic display, on the other hand, is a true $3$D display technology, which can provide all depth cues such as binocular parallax, motion parallax, accommodation and convergence, without having to wear any special glasses. It promises no eye fatigue and has big market potential for various applications for entertainment, advertisement, medical imaging, training and education, defence and military, and scientific visualisation.

A spatial light modulator (SLM) is an optical device that can spatially modulate the amplitude (or intensity), phase or polarisation of an optical wave. SLM is a key device for digital $3$D holographic display \cite{Xu10}. The pixel pitch and pixel count are two key SLM parameters that affect both the size and viewing angle of reconstructed $3$D images. For practical applications of the $3$D holographic display technology, advanced SLM devices with submicron resolution and ultrafast response time need to be developed and manufactured with high yield. However, the current commercial technologies do not meet these requirements.

Consequently, Aoshima et. al. \cite{Aos10} proposed and demonstrated a spin-transfer-switching-based \cite{Sta14} SLM device that exploits the magneto-optical spatial light-modulation effect and is driven by spin-polarised current flow. The proposed SLM device has a spatial resolution as small as several hundred nanometers and possesses the potential for being driven at ultra-high speed of several tens of nanoseconds. However, even these characteristics just marginally satisfy both the size and speed requirements of SLMs for use in displaying holographic $3$D moving images. We anticipate that the plasmon-enhanced magneto-optical interaction together with the possibility to focus plasmons into sub-$100$ nm spots may be used to further improve the performance of this SLM device. 

Finally, we would like to mention that plasmon-enhanced magneto-optical effects may help to improve the efficiency of portable medical devices for clinical and in-field tests. For instance, malaria is a deadly infection caused by species of the Plasmodium parasite, passed on via the bite of an infected mosquito. Diagnosis is made by a blood test. However, sometimes it is necessary to repeat the test a number of times, as the parasites can be difficult to detect. A new method proposed in Ref. \cite{Orban} exploits magneto-optical effects and it allows high-sensitivity detection of malaria pigment (hemozoin crystals) in blood via the magnetically induced rotational motion of the hemozoin crystals. We believe that the plasmon-enhanced magneto-optical interaction may further improve the sensitivity of the proposed method.   
  
\section{Conclusions}

We reviewed several important directions in the study of magneto-plasmonics and resonant light-magnetic-matter interaction at the nanoscale. We endeavoured to present the material from two different points of view: magneto-plasmonics, and magnonics and magnetisation dynamics. Our main aim was to try to combine research outcomes produced by both research sectors, and we also tried to focus on novel results and/or results that were not highlighted in the previous review papers. It is clear that this research area is yet to reach its peak and to provide a valuable overall impact on the fields of photonics and nanomagnetism.


\acknowledgments{Acknowledgements}

The author has been supported by the UPRF scheme of the University of Western Australia. He is very grateful to Prof. M. Kostylev (UWA) for careful reading of the manuscript and valuable suggestions. He also thanks N. Kostylev, S. Samarin, J. Williams, A. Adeyeye, P. Metaxas, L. Willig, and J. Hutomo for useful discussions.

\end{document}